\documentclass[aps,prb,reprint,showpacs]{revtex4-1}
\usepackage{graphicx}

\begin{document}

\title{Band offsets in heterojunctions between cubic perovskite oxides}

\author{Alexander I. Lebedev}
\email[]{swan@scon155.phys.msu.ru}
\affiliation{Physics Department, Moscow State University,\\
Leninskie gory, 119991 Moscow, Russia}

\date{\today}

\begin{abstract}
The band offsets for nine heterojunctions between titanates, zirconates,
and niobates with the cubic perovskite structure were calculated from first
principles. The effect of strain in contacting oxides on their energy
structure, many-body corrections to the position of the band edges
(calculated in the $GW$ approximation), and the splitting of the conduction
band resulting from spin-orbit interaction were consistently taken into
account. It was shown that the neglect of the many-body effects can lead
to errors in determination of the band offsets up to 0.36~eV. The failure of
the transitivity rule, which is often used to determine the band offsets in
heterojunctions, was demonstrated and its cause was explained.
\end{abstract}


\maketitle

\section{Introduction}

Almost all electronic and optoelectronic devices contain
metal--semiconductor, metal--dielectric, semiconductor--dielectric, or
semiconductor--semiconductor interfaces. As the energy of an electron abruptly
changes at the interface, the characteristics of devices that include such
interfaces directly depend on the height of emerging energy barriers.
Originally, the concept of a heterojunction was associated with a contact
between two semiconductors, but currently its use is significantly expanded
to include dielectrics. For example, in solving the actual problem of
replacing the SiO$_2$ gate dielectric in silicon field-effect transistors
with a material with higher dielectric constant, the calculation of the
tunneling current through the gate dielectric requires an accurate knowledge
of the energy diagram of the formed heterojunction.

In the last decade, experimental studies have discovered a number of new
physical phenomena occurring at the interface of two oxide dielectrics.
These include the appearance of a quasi-two-dimensional electron gas (2DEG)
at the interface of two dielectrics~\cite{Nature.427.423} and its
superconductivity;~\cite{Science.317.1196}  the magnetism at the interface
of two non-magnetic oxides.~\cite{NatureMater.6.493}  The ability to control
the conductivity of the 2DEG with an electric field (an analog of the field
effect)~\cite{Science.313.1942,NatureMater.7.298}  and so to control the
temperature of the superconducting phase transition~\cite{Nature.456.624}
were demonstrated. The strongest conductivity changes were observed when one
of the oxides was ferroelectric.~\cite{Science.284.1152}  When the components
of heterostructures were magnetic and ferroelectric oxides, the ability to
control the magnetic properties of the structure with a switchable ferroelectric
polarization was shown.~\cite{PhysRevB.68.134415,ApplPhysLett.89.242506,
AdvMater.21.3470} Thus, these heterostructures acquired the properties of a
multiferroic. The above-mentioned and other interesting phenomena observed in
oxide heterostructures open new opportunities for developing of new
multifunctional electronic devices and suggest the emergence of a new direction
in microelectronics, the oxide electronics.~\cite{Science.327.1607,
AnnuRevCondensMatterPhys.2.141,NatureMater.11.103}

The development of the ferroelectric memory devices is one of the applications
of the ferroelectric oxides. It requires to solve the problem of non-destructive
read-out of information and to increase the packing density of the memory cells.
For non-destructive optical read-out methods, the memory cell size is limited
by the wave length. For titanates with the perovskite structure, in which the
typical band gap is $\sim$3~eV, the minimum cell size is $\sim$4000~{\AA}. In
multiferroics, in which the information is stored electrically and read out
magnetically, the memory cell size can be reduced to the size typical for modern
hard disks, $\sim$500~{\AA} (in homogeneous multiferroic thin film, the physical
size of the memory cells is limited by a rather large width of magnetic domain
wall).

The methods based on the electrical read-out of the ferroelectric polarization
are the most promising. Since the physical size of the ferroelectric memory
cells is limited by the thickness of the ferroelectric domain wall and the
minimum film thickness for which the ferroelectricity still exists (both sizes
are a few unit cells~\cite{PhysRevB.53.R5969,PhysRevB.65.104111,Nature.422.506,
PhysRevB.72.020101}), the packing density of the memory cells in these devices
is maximized. These methods can be based, for example, on the nonlinear
current-voltage characteristics which are reversible upon switching the
polarization in the metal--ferroelectric--metal structures as demonstrated on
thin films of PZT~\cite{ApplPhysLett.83.4595} and BiFeO$_3$.~\cite{Science.324.63,
ApplPhysLett.98.192901}  A similar behavior can be observed on the structures
that use the tunneling through a thin layer of
ferroelectric.~\cite{PhysRevLett.94.246802,PhysRevB.72.125341,Nature.460.81}

The most important physical parameters that characterize an interface between
two semiconductors or dielectrics are the band offsets on the energy diagram
of the heterojunction. The valence band offset $\Delta E_v$ (the conduction
band offset $\Delta E_c$) is defined as the difference between the energies of
the tops of the valence bands (bottoms of the conduction bands) in two contacting
materials. These band offsets determine a number of physical properties of
heterojunctions, in particular, their electrical and optical properties.

For oxides with the perovskite structure, the reliable experimental data
on the band offsets (obtained by the photoelectron spectroscopy) exist for
heterojunctions SrTiO$_3$/Si,~\cite{ApplPhysLett.77.1662,JVacSciTechnolA.19.934,
JApplPhys.96.1635}
BaTiO$_3$/Si,~\cite{JApplPhys.96.1635}
SrTiO$_3$/GaAs,~\cite{ApplPhysLett.86.082905,ApplPhysLett.103.031919}
BaTiO$_3$/Ge,~\cite{JApplPhys.114.024303}
SrTiO$_3$/InN,~\cite{NanoscaleResLett.6.193,InTech.Ferr.16}
BaTiO$_3$/InN,~\cite{InTech.Ferr.16}
SrTiO$_3$/ZnO,~\cite{InTech.Ferr.16}
BaTiO$_3$/ZnO,~\cite{ApplPhysA.99.511,InTech.Ferr.16}
SrTiO$_3$/SrO,~\cite{PhysRevB.67.155327}
and BaTiO$_3$/BaO.~\cite{PhysRevB.67.155327}
For heterojunctions between two perovskite dielectrics, the data are
limited by the PbTiO$_3$/SrTiO$_3$,~\cite{PhysRevB.84.045317}
SrTiO$_3$/SrZrO$_3$,~\cite{JPhysD.45.055303}
SrTiO$_3$/LaAlO$_3$,~\cite{JPhysCondensMatter.22.312201,PhysRevB.88.115111}
and SrTiO$_3$/BiFeO$_3$~\cite{NewJPhys.15.053014} systems. In addition, there
are data on the Schottky barrier heights for perovskite--metal structures,
in which the metals are Pt, Au, Ag, or the conductive oxides SrRuO$_3$ and
(La,Sr)CoO$_3$.

In this work, the band offsets for nine heterojunctions between titanates,
zirconates, and niobates with the cubic perovskite structure are calculated from
first principles using the density functional theory and $GW$ approximation.
The obtained results are compared with available experimental data.

\section{Methodology}

The band offsets cannot be determined from a direct comparison of the
energies of the valence and conduction bands obtained from the first-principles
band-structure calculations performed separately for two constituent bulk
materials. This is because in the first-principles calculations, there is no
intrinsic energy scale: the energies corresponding to the valence band edge
$E_v$ and to the conduction band edge $E_c$ are usually measured from an
average of the electrostatic potential, which is an ill-defined quantity for
infinite systems. Consequently, in addition to the band-structure calculations
for two materials, the lineup of the average of the electrostatic potential
$\Delta V$ between them should also be calculated. The latter value is
determined by the dipole moment emerging at the heterojunction as a result
of the redistribution of the electron density on the hybridized orbitals in
the constituent materials. It takes into account all the features typical
of the interface, such as the change in the chemical composition, the structure
distortions, etc.

Thus, the valence band offset can be represented as a sum of two
terms:~\cite{PhysRevB.44.5572}
\begin{equation}
\Delta E_v = (E_{v2} - E_{v1}) + \Delta V.
\label{eq1}
\end{equation}

The first term in this equation is the difference of the energies corresponding
to the tops of the valence bands in two bulk materials. It can be obtained
from standard band-structure calculations.

The second term in Eq.~(\ref{eq1}) is the lineup of the average of the
electrostatic potential through the heterojunction.
To calculate $\Delta V$, one usually starts from the total potential (the
potential of the ions plus the microscopic electrostatic Hartree potential for
electrons) obtained from the self-consistent electron density calculation in
a superlattice constructed of the contacting materials. After that, the
macroscopic averaging technique~\cite{PhysRevLett.61.734}  is applied, in
which the electrostatic potential is first averaged over planes parallel to
the interface and then the convolution of the obtained one-dimensional
quasi-periodic function with two rectangular windows whose lengths are
determined by the periods of components is calculated. The resulting profiles
of the macroscopic average of the electrostatic potential $\bar{V}(r)$ have
flat (bulk-like) regions far enough from the interface. The $\Delta V$ value
is defined as the difference energy between these plateau values. It should
be noted that neither the $E_{v1}$ and $E_{v2}$ quantities, nor $\Delta V$
does not have physical meaning themselves, only the sum, Eq.~(\ref{eq1}), is
meaningful.

The conduction band offset is calculated from the $\Delta E_v$ value and
the difference of the band gaps in two materials:
$$\Delta E_c = (E_{c2} - E_{c1}) + \Delta V = (E_{g2} - E_{g1}) + \Delta E_v.$$

Roughly, the band gap $E_g = E_c - E_v$ can be estimated in the LDA
approximation for the exchange-correlation energy. However, because of the
well-known band-gap problem characteristic of this one-electron
approximation, more accurate calculations should take into account the
corrections to the band energies resulting from many-body effects. These
corrections (the values $\Delta E_c^{\rm QP}$ and $\Delta E_v^{\rm QP}$) are
usually calculated in the quasiparticle $GW$ approximation. It is usually
believed that many-body corrections adjust the position of the conduction
band and in this way solve the band-gap problem, however, the energy levels
in the valence band are also corrected.

In the case of well-studied materials, the $\Delta E_c$ value is often
calculated from the experimental band gaps. However, if the band offset
$\Delta E_v$ was obtained theoretically, the problem associated with
the uncertainty in the $\Delta E_v^{\rm QP}$ values remains. It is often
assumed that the $\Delta E_v^{\rm QP}$ values in two materials are close,
so that their contributions to the band offsets cancel each other. In
this work, we show that this assumption in general is not true.

It should be also noted that, since the position of the energy levels in
a crystal depends on the interatomic distances, the calculation of $E_{v1}$,
$E_{c1}$, $E_{v2}$, and $E_{c2}$ should be carried out under the same strain
of materials, which appears in a heterojunction. Besides the strain-induced
lifting of the degeneracy of the band edges, the band gap itself can vary.
Moreover, the calculations should take into account the possible lifting of
the band degeneracy resulting from spin-orbit interaction. Although the
structure of dielectrics---the lattice parameters and equilibrium atomic
positions---are weakly dependent on the spin-orbit interaction (and therefore
it is usually neglected in the calculations), in the band structure
calculations, the spin-orbit interaction strongly affects the energy position
of the bands, and it cannot be neglected. In dielectrics, the taking into
account of the spin-orbit interaction can be done \emph{a posteriori}, after
completion of the main first-principles calculations.

\section{The calculation technique}

The objects of the present calculations were heterojunctions formed between
titanates and zirconates of calcium, strontium, barium, and lead as well as
the KNbO$_3$/NaNbO$_3$ heterojunction. They were modeled using superlattices
grown in the [001] direction and consisted of two materials with equal
thickness of layers, each of four unit cells. The in-plane lattice parameter
was obtained from the condition of zero stress at the interface (i.e., it
was close to the lattice parameter of the solid solution with the component
ratio of 1:1); the period of the superlattice and atomic displacements normal
to the interface were fully relaxed.

The equilibrium lattice parameters and atomic positions were calculated from
first principles within the density functional theory (DFT) using the
\texttt{ABINIT} software. The exchange-correlation interaction was described
in the local density approximation (LDA). Pseudopotentials of the atoms were
taken from Refs.~\onlinecite{PhysSolidState.51.362,PhysSolidState.52.1448}.
The maximum energy of plane waves was 30~Ha (816~eV). For the integration over
the Brillouin zone, the 8$\times$8$\times$2 Monkhorst-Pack mesh was used.
All calculations were performed for heterojunctions between cubic $Pm3m$ phases;
the possible polar and antiferrodistortive (structural) distortions of the
materials were neglected (they will be considered in a separate paper). The
$\Delta V$ value was determined using the macroscopic averaging
technique.~\cite{PhysRevLett.61.734}  To determine the values of $E_{v1}$,
$E_{c1}$, $E_{v2}$, and $E_{c2}$ in constituent materials, similar calculations
were performed for isolated crystals with the in-plane lattice parameter equal
to that of the superlattice; in the normal direction the crystals were
stress-free.

The calculations of the quasiparticle band gap and the many-body corrections
to the position of the band edges were carried out in the so-called one-shot
$GW$ approximation.~\cite{RevModPhys.74.601}  The Kohn-Sham wave functions and
energies calculated within the DFT-LDA approach were used as a zeroth-order
approximation. The dielectric matrix
$\epsilon_{\mathbf{GG'}}(\mathbf{q},\omega)$ was calculated for
a 6$\times$6$\times$6 mesh of wave vectors $\mathbf{q}$ from the matrix of
irreducible polarizability
$P^0_{\mathbf{GG'}}(\mathbf{q},\omega)$ calculated for 2200--2800 vectors
$\mathbf{G}(\mathbf{G'})$ in reciprocal space, 20--22 filled and 278--280
empty bands. Dynamic screening was described in the Godby--Needs plasmon-pole
model. The wave functions with the energies up to 24 Ha were taken into account
in the calculations. The energy corrections to the LDA solution were calculated
from the diagonal matrix elements of [$\Sigma - E_{xc}$] operator, where
$\Sigma = GW$ is the self energy operator, $E_{xc}$ is the operator of the
exchange-correlation energy, $G$ is the Green's function, and
$W = \epsilon^{-1}v$ is the screened Coulomb interaction. In calculating
$\Sigma$, the wave function with the energies up to 24~Ha were taken into
account.

\section{Results}

\begin{figure}
\centering
\includegraphics{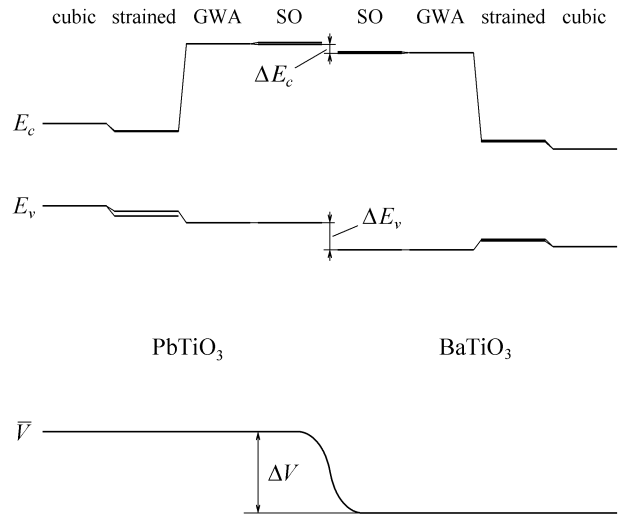}
\caption{\label{fig1}The steps of the band offsets calculation for the
PbTiO$_3$/BaTiO$_3$ heterojunction. First, the changes in the band structure
of cubic phases induced by the strain in the heterojunction are taken into
account, then the corrections resulting from the many-body effects are applied
(GWA), and finally, the splitting of the band edges caused by the spin-orbit
interaction (SO) is taken into account. The bottom chart shows the change of
the average of the electrostatic potential in the heterojunction, which is a
reference energy level for all bands.}
\end{figure}

\begin{figure}
\centering
\includegraphics{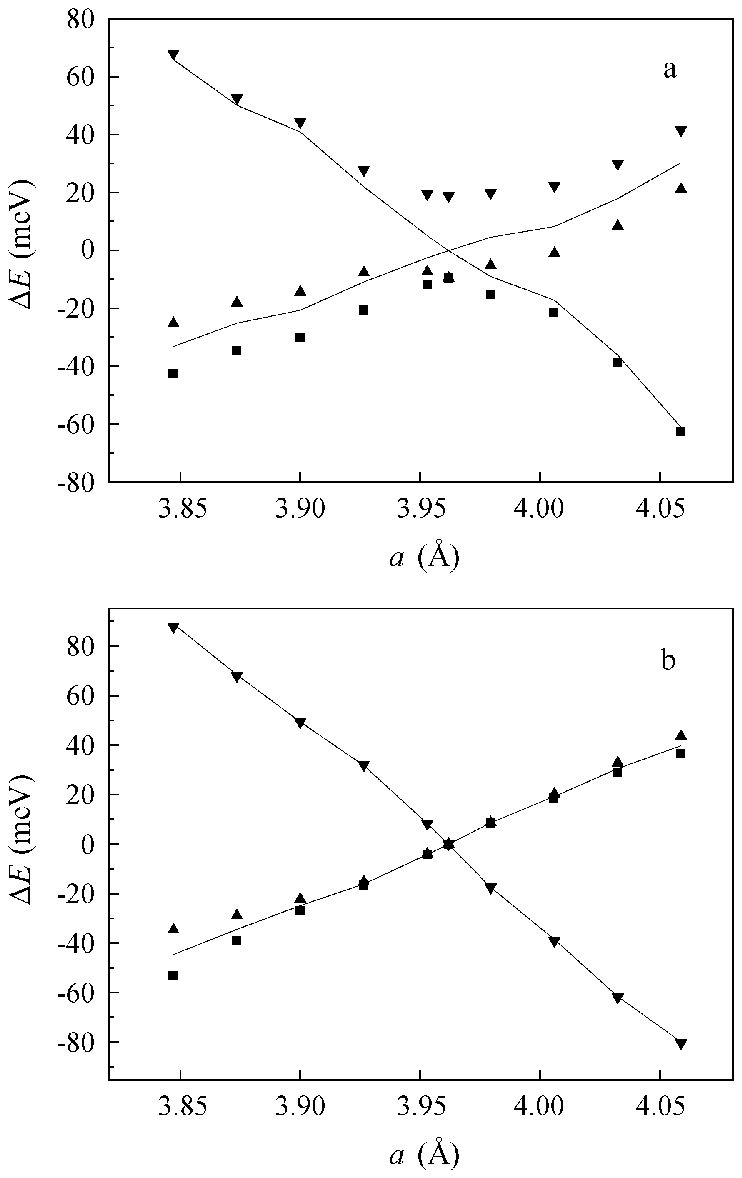}
\caption{\label{fig2}Effect of biaxial strain on the splitting of the conduction
band (a) and of the valence band (b) in BaTiO$_3$ without the spin-orbit
interaction (lines) and with it (dots). The lattice parameter of stress-free
(cubic) crystal is 3.962~{\AA}.}
\end{figure}

Fig.~\ref{fig1} shows the steps of the energy diagram calculation for a typical
heterojunction. The extreme left and right parts in the figure correspond
to the individual compounds with the cubic $Pm3m$ structure, whose lattice
parameter corresponds to zero external stress. Biaxial strain of these
compounds during the formation of the heterojunction (their in-plane lattice
parameters become equal), reduces the symmetry of their unit cells to $P4/mmm$.
As a result, their band gaps are changed and the degeneracy of the energy levels
at some points of the Brillouin zone are lifted. For example, the threefold
degeneracy of the conduction band at the $\Gamma$~point and that of the valence
band at the $R$~point are lifted (Fig.~\ref{fig2}). These band extrema locations
are typical for all compounds studied in this work except for PbTiO$_3$
and PbZrO$_3$. In cubic PbTiO$_3$, the top of the valence band is at the
$X$~point, and the tetragonal distortion removes the valley degeneracy (depending
on the sign of the strain, the valence band edge is located at either $X$ or
$Z$~point of the Brillouin zone of the tetragonal lattice). In cubic PbZrO$_3$,
the only compound in which both band extrema are located at the $X$~point,
the strain also removes the valley degeneracy, but both extrema remain at
the same point of the Brillouin zone ($X$ or $Z$). The energy diagrams of
strained crystals are shown in Fig.~\ref{fig1} next to the diagrams for cubic
phases. We note that not only the band gap, but also the energies of the band
edges ($E_c$ and $E_v$) measured from the averaged electrostatic potential are
changed upon strain. The values of these energies in strained crystals are given
in Tables~\ref{table1} and \ref{table2}.

\begin{table*}
\caption{\label{table1}Parameters determining the valence band offset
$\Delta E_v$ on the energy diagram of heterojunctions (all the energies are
in eV).}
\begin{ruledtabular}
\begin{tabular}{ccccccc}
Heterojunction   & $E_{v2}$ & $\Delta E_{v2}^{\rm QP}$ & $E_{v1}$ & $\Delta E_{v1}^{\rm QP}$ & $\Delta V$ & $\Delta E_v$ \\
\hline
SrTiO$_3$/PbTiO$_3$ & 13.629  & $-$0.239 & 15.464  & $-$0.315  &   +2.143 &   +0.384 \\
BaTiO$_3$/BaZrO$_3$ & 13.422  & $-$0.512 & 13.766  & $-$0.226  &   +0.066 & $-$0.564 \\
PbTiO$_3$/PbZrO$_3$ & 12.390  & $-$0.321 & 13.123  & $-$0.239  &   +0.495 & $-$0.320 \\
PbTiO$_3$/BaTiO$_3$ & 14.291  & $-$0.226 & 13.453  & $-$0.239  & $-$1.276 & $-$0.425 \\
SrTiO$_3$/BaTiO$_3$ & 14.366  & $-$0.226 & 15.333  & $-$0.315  &   +0.864 & $-$0.014 \\
SrTiO$_3$/SrZrO$_3$ & 14.391  & $-$0.582 & 14.912  & $-$0.315  &   +0.395 & $-$0.393 \\
PbZrO$_3$/BaZrO$_3$ & 13.158  & $-$0.512 & 11.888  & $-$0.321  & $-$1.209 & $-$0.130 \\
SrTiO$_3$/CaTiO$_3$ & 15.631  & $-$0.333 & 15.664  & $-$0.315  &   +0.131 &   +0.080 \\
KNbO$_3$/NaNbO$_3$  & 13.617  & $-$0.314 & 14.494  & $-$0.245  &   +0.944 & $-$0.002 \\
\end{tabular}
\end{ruledtabular}
\end{table*}

\begin{table*}
\caption{\label{table2}Parameters determining the conduction band offset
$\Delta E_c$ on the energy diagram of heterojunctions and the type of the
heterojunction (all the energies are in eV).}
\begin{ruledtabular}
\begin{tabular}{ccccccccc}
Heterojunction     & $E_{c2}$ & $\Delta E_{c2}^{\rm QP}$ & $\Delta E_{c2}^{\rm SO}$ & $E_{c1}$ & $\Delta E_{c1}^{\rm QP}$ & $\Delta E_{c1}^{\rm SO}$ & $\Delta E_c$ & Type \\
\hline
SrTiO$_3$/PbTiO$_3$ & 14.899  & +1.326 & $-$0.010 & 17.034  & +1.431 & $-$0.007 & $-$0.100 & I \\
BaTiO$_3$/BaZrO$_3$ & 16.363  & +1.199 & $-$0.026 & 15.143  & +1.341 & $-$0.008 &   +1.126 & I \\
PbTiO$_3$/PbZrO$_3$ & 14.513  & +0.266 &    0     & 14.269  & +1.326 & $-$0.010 & $-$0.311 & II \\
PbTiO$_3$/BaTiO$_3$ & 15.820  & +1.341 & $-$0.008 & 14.704  & +1.326 & $-$0.010 & $-$0.143 & II \\
SrTiO$_3$/BaTiO$_3$ & 15.885  & +1.341 & $-$0.008 & 16.879  & +1.431 & $-$0.007 & $-$0.221 & II \\
SrTiO$_3$/SrZrO$_3$ & 17.469  & +1.283 & $-$0.023 & 16.347  & +1.431 & $-$0.007 &   +1.353 & I \\
PbZrO$_3$/BaZrO$_3$ & 16.116  & +1.199 & $-$0.026 & 14.069  & +0.266 &    0     &   +1.745 & I \\
SrTiO$_3$/CaTiO$_3$ & 17.207  & +1.486 & $-$0.007 & 17.237  & +1.431 & $-$0.007 &   +0.156 & II \\
KNbO$_3$/NaNbO$_3$  & 15.005  & +1.008 & $-$0.038 & 15.823  & +0.976 & $-$0.037 &   +0.157 & I \\
\end{tabular}
\end{ruledtabular}
\end{table*}

The calculation of the many-body corrections to the positions of the valence
band edge $\Delta E_v^{\rm QP}$ and of the conduction band edge
$\Delta E_c^{\rm QP}$ in the $GW$ approximation shows that the many-body
effects shift the position of the conduction band up by $\sim$1.3~eV in all
compounds studied in this work except for PbZrO$_3$ in which the shift is
only 0.266~eV (Tables~\ref{table1} and \ref{table2}). The valence band edge
is shifted down by 0.22--0.58~eV on taking into account the many-body effects.
Although the absolute values of these corrections slowly converge with
increasing number of empty bands in the $GW$ calculations (see, for example,
Ref.~\onlinecite{PhysRevB.78.085125}), the relative shift of the difference
between the corrections in different compounds is small. Therefore, if one
uses the same total number of bands (300 in our case) in the calculations of
these corrections, the error in determination of the relative position of the
band edges in two materials is not large, $\sim$0.01~eV according to our
estimates. In our calculations we have also assumed that the many-body
corrections slightly depend on the strain, and so the values calculated for the
cubic crystals can be used. The tests have shown that the strain-induced changes
of $\Delta E_v^{\rm QP}$ and $\Delta E_c^{\rm QP}$ may reach 0.01--0.02~eV,
which gives an estimate of possible errors. The energy diagrams of constituent
materials after accounting for many-body effects are also shown in
Fig.~\ref{fig1}.

The calculations of many-body corrections show that the assumption used
by many authors about an approximate equality of these corrections
in two contacting materials, in general, is not true. It is seen that in
related oxides with the cubic perovskite structure the scatter of the
$\Delta E_v^{\rm QP}$ values can be as large as 0.36~eV. This value is a
measure of the possible error in determination of the band offsets in the
calculations that neglect the many-body effects.%
    \footnote{A more detailed study have shown that in a large class of oxides,
    fluorides, and nitrides the variation in the $\Delta E_v^{\rm QP}$ values
    can reach 3~eV. These results and their explanation will be published in
    a separate paper.}

Since our crystals contain atoms with a large atomic number, the errors
in determination of the band edge positions resulting from the neglect of
the spin-orbit interaction can be quite large. In this work, the spin-orbit
splitting $\Delta_{\rm SO}$ of the valence and conduction band edges was
calculated using the full-relativistic pseudopotentials.~\cite{PhysRevB.58.3641}
The tests performed on a number of semiconductors (Ge, GaAs, CdTe), for
which the spin-orbit splitting of the valence band is accurately measured,
showed that the $\Delta_{\rm SO}$ values calculated in this way agree with
experiment to within $\sim$5\%.

The calculations showed that the spin-orbit interaction results in
a splitting of the band edges at some points of the Brillouin zone. First
of all, it refers to the conduction band edge at the $\Gamma$ point. It is
interesting that despite the presence of heavy atoms such as Ba and Pb in
our crystals, the spin-orbit splitting is not very large. This is because
the conduction band states at the $\Gamma$~point in these crystals are
formed primarily from the $d$-states of the $B$~atom (Ti, Zr, Nb). The
values $\Delta_{\rm SO}$ of the spin-orbit splitting of the conduction
band at the $\Gamma$~point for all materials studied except for PbZrO$_3$
are given in Table~\ref{table3}. In PbZrO$_3$, the conduction band minimum
is located at the $X$~point, is non-degenerate, and does not exhibit
the spin-orbit splitting. The valence band edge (at $R$ and $X$ points) in
all cubic crystals studied in this work does not exhibit the spin-orbit
splitting too.

\begin{table*}
\caption{\label{table3}The spin-orbit splitting $\Delta_{\rm SO}$ of the
conduction band at the $\Gamma$~point in cubic perovskites (in meV).}
\begin{ruledtabular}
\begin{tabular}{cccccccc}
CaTiO$_3$ & SrTiO$_3$ & BaTiO$_3$ & PbTiO$_3$ & SrZrO$_3$ & BaZrO$_3$ & NaNbO$_3$ & KNbO$_3$ \\
\hline
20.7      & 22.0      & 25.3      & 28.5      & 70.2      & 77.5      & 113.7     & 111.0 \\
\end{tabular}
\end{ruledtabular}
\end{table*}

When the spin-orbit interaction is turned on, the center of gravity of the
split energy levels coincides with the position of the energy level without
spin-orbit interaction.~\cite{PhysRevB.34.5621} In all studied crystals, the
spin-orbit split-off conduction band at the $\Gamma$~point is always shifted
to higher energies, and so the absolute minimum of the conduction band at the
$\Gamma$~point is shifted down by $\Delta_{\rm SO}/3$. This value is given
in Table~\ref{table2} and determines an additional shift of the conduction
band. The final energy diagram of the heterojunction obtained after taking
into account the spin-orbit interaction is shown by two internal diagrams in
Fig.~\ref{fig1}. We note that in these calculations, we have neglected the
weaker effects associated with the change of the band splitting; they result
from the strain-induced mixing of different states and can be seen in
Fig.~\ref{fig2}. These effects, however, do not exceed 10~meV and are less
than other systematic errors in our calculations.

\begin{figure}
\centering
\includegraphics{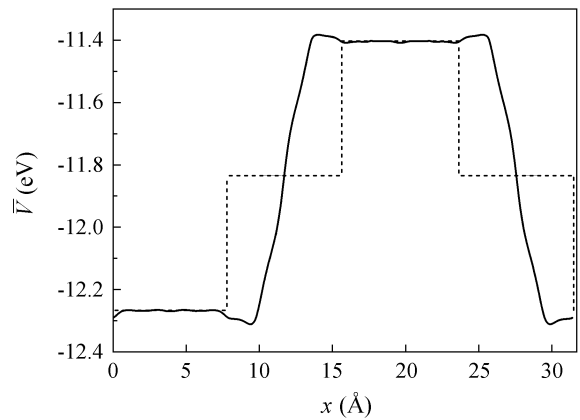}
\caption{\label{fig3}Determination of the $\Delta V$ form the profile of the
average electrostatic potential $\bar{V}(x)$ for the SrTiO$_3$/BaTiO$_3$
superlattice (solid line). The dotted line shows the approximating function.}
\end{figure}

In calculating the $\Delta V$ value, the profile of the electrostatic potential
$\bar{V}(x)$ obtained with the macroscopic averaging technique was approximated
by a step function with a width of transition regions of one lattice parameter
(Fig.~\ref{fig3}). The tests showed that when the thickness of the individual
layers in the BaTiO$_3$/SrTiO$_3$ superlattice was changed from three to five
unit cells, the variation in the $\Delta V$ value computed by the above
algorithm was only $\sim$4~meV, which gives an estimate of the error in the
$\Delta V$ calculation. As shown in Ref.~\onlinecite{PhysRevLett.100.186401},
the many-body effects do not influence much on the $\Delta V$ value.

\begin{figure*}
\centering
\includegraphics[scale=1.5]{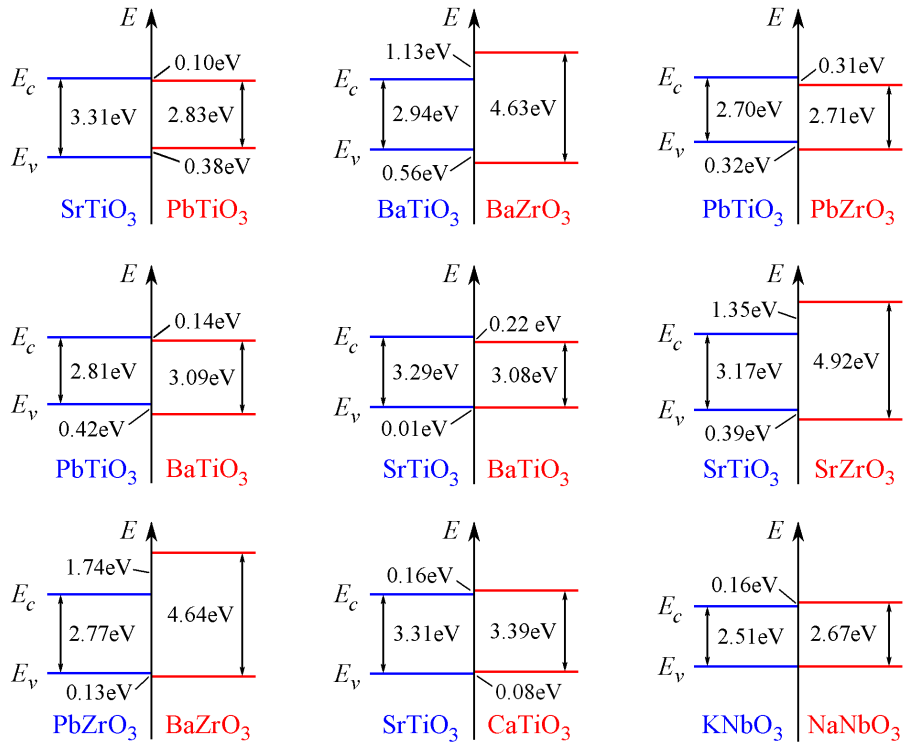}
\caption{(Color online) Energy diagrams for heterojunctions between cubic
perovskites studied in this work.}
\label{fig4}
\end{figure*}

The results of the band offsets calculations for nine heterojunctions are
given in Tables~\ref{table1} and~\ref{table2}. The signs of the band
offsets are defined as the energy change when moving from the compound
indicated the first in the heterojunction pair to the compound indicated the
second. The energy diagrams of heterojunctions are classified as type~I
heterojunctions, for which $\Delta E_c$ and $\Delta E_v$ have opposite signs,
and type~II heterojunctions, for which the signs of $\Delta E_c$ and
$\Delta E_v$ are the same. The types of the heterojunctions are given in
Table~\ref{table2} and their energy diagrams are shown in Fig.~\ref{fig4}.

\section{Discussion}

Unfortunately, the experimental data on the band offsets in heterojunctions
between oxides with the perovskite structure are very limited. In
Ref.~\onlinecite{JPhysD.45.055303}, the band offsets for the
SrTiO$_3$/SrZrO$_3$ heterojunction were obtained using the photoelectron
spectroscopy. According to these measurements, the heterojunction is type~I,
and the band offsets are $\Delta E_v = -0.5 \pm 0.15$~eV and
$\Delta E_c = +1.9 \pm 0.15$~eV (in SrTiO$_3$ the valence band lies higher
than in SrZrO$_3$). Our data agree well with these experimental data:
according to our calculations, the heterojunction is also type~I, and the
band offsets are $-$0.393 and +1.353~eV, respectively, for \emph{cubic}
components.%
    \footnote{If we take into account that the real structure of SrZrO$_3$
    at 300~K is orthorhombic and the corresponding increase of the LDA
    band gap in this structure is 0.361~eV, the conduction band offset can
    be estimated as +1.714~eV. In this estimate, we neglected the changes of
    the many-body corrections upon structural distortion and supposed that
    the band gap correction changes only the location of the conduction band.}
We note that the band
offsets predicted in this work are in much closer agreement with experiment
compared to the results of Ref.~\onlinecite{PhysRevB.88.115304}
($\Delta E_v = +0.5$~eV, $\Delta E_c = +2.5$~eV).

As the experimental SrTiO$_3$/SrZrO$_3$ structure~\cite{JPhysD.45.055303} was
grown on SrTiO$_3$ substrate, the calculated band offsets, which depend on the
in-plane lattice parameter, may be slightly different (this dependence is well
known for semiconductor heterojunctions~\cite{PhysRevB.34.5621,PhysRevB.44.5572,
SolidStateCommun.124.63}). To estimate these changes, the calculations for the
SrTiO$_3$/SrZrO$_3$ heterojunction were repeated for the in-plane lattice
parameter equal to that of SrTiO$_3$. They gave $\Delta E_v = -0.240$~eV and
$\Delta E_c = +1.230 $~eV, which somewhat worsened the agreement with
experiment.

It should be stressed that the results obtained in this work are for
heterojunctions formed between \emph{cubic} compounds. We deliberately did
not take into account possible distortions of the perovskite structure, which
obviously affect the energy diagram of heterojunctions. This is because the
question about the character of these distortions is not so simple. It is
known that the character of distortions in the perovskites may change
significantly under the biaxial strain and, moreover, the distortions in two
materials usually are highly interconnected. These effects are well known for
ferroelectric superlattices.~\cite{PhysRevB.71.100103, PhysRevB.72.214121,
PhysSolidState.51.2324, PhysSolidState.52.1448}  In heterojunctions between
polar materials, the need to meet the electrical boundary conditions (equal
electric displacement fields normal to the interface) causes the polarization
in both constituent materials to differ from their equilibrium values. Since
the accompanying atomic displacements affect the band gap and the positions
of the band edges, in polar heterojunctions the band offsets can be very
different from those in nonpolar structures.%
    \footnote{The same reasoning applies to heterojunctions between
    semiconductors with the wurtzite structure like GaN/AlN in which both
    materials have nonzero spontaneous polarization.}
In addition, the cases are known when the periodic domain structure can
occur in a ferroelectric near the interface.~\cite{PhysRevB.82.045426}
Predicting of the energy diagram for such a system is particularly problematic.

The experimental data for the SrTiO$_3$/PbTiO$_3$
heterojunction~\cite{PhysRevB.84.045317} are noticeably different from the
results of our calculations. According to the photoelectron spectroscopy
data, this heterojunction is type~II and the band offsets are
$\Delta E_v = +1.1 \pm 0.1$~eV and $\Delta E_c = +1.3 \pm 0.1$~eV (in PbTiO$_3$
the valence band lies higher than in SrTiO$_3$). According to our calculations,
the band offsets are +0.384 and $-$0.100~eV, respectively, and the
heterojunction is type~I. We see that in experiment and calculations,
the signs of $\Delta E_v$ are the same, but their values differ considerably.
The fact that the crystal structure of PbTiO$_3$ at 300~K is tetragonal cannot
explain such a large discrepancy. A possible explanation is given below.

If the interface is not perfect (for example, in the case when the structural
relaxation of strained materials occurs, as one can see in
Refs.~\onlinecite{JPhysD.45.055303, NewJPhys.15.053014}), the dangling bonds
are created near the interface and the surface states appear in the electronic
structure. These states are electrically active and can significantly disturb
the $\Delta V$ value, and so to change $\Delta E_c$ and $\Delta E_v$.
In addition, in the relaxation region where the lattice parameter depends on
the coordinate, an additional drift of the $E_v$ and $E_c$ values (the band
bending) occurs. The distortion of the energy diagrams similar to that
arising from the surface states can occur in heterojunctions between highly
defective materials. Like the levels of the surface states, defects in
constituent materials can exchange electrons with each other, which would
distort the energy diagram of a heterojunction. The size of the region in
which such an exchange occurs is about the screening length in a material and
can be quite small. For example, in a dielectric with a defect concentration
of $10^{18}$~cm$^{-3}$, the length of the impurity screening can be as small as
43~{\AA}.~\cite{PhysRev.164.1025}  It is possible that the strong discrepancy
between the calculated and experimental band offsets in the SrTiO$_3$/PbTiO$_3$
heterojunction results from the presence of defects in materials: the band
offsets observed in this heterojunction correspond to the case where the defect
levels in two contacting materials are close in energy.~\cite{PhysRevB.84.045317}
Even stronger distortions of the band diagram can be characteristic of
heterostructures between perovskites with so-called valence discontinuity, such
as SrTiO$_3$/LaAlO$_3$ (Ref.~\onlinecite{JPhysCondensMatter.22.312201,
PhysRevB.88.115111}) and BiFeO$_3$/SrTiO$_3$ (Ref.~\onlinecite{NewJPhys.15.053014}),
in which the distribution of the quasi-two-dimensional electron gas density is
different from that of the positive (ionic) charge.

\begin{figure}
\centering
\includegraphics{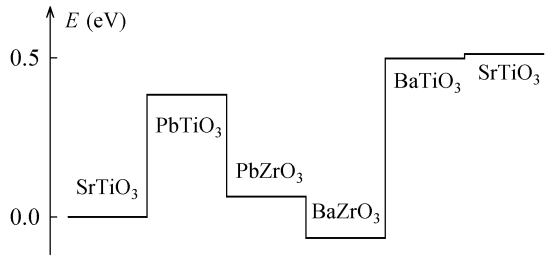}
\caption{\label{fig5}Changes of the valence band position when traversing
the contour SrTiO$_3$/PbTiO$_3$/PbZrO$_3$/BaZrO$_3$/BaTiO$_3$/SrTiO$_3$.}
\end{figure}

In conclusion, we discuss the applicability of the transitivity rule which is
often used to calculate the band offsets in heterojunctions by comparing the
band offsets in a pair of heterojunctions formed between the components of
the heterojunction under discussion and the third common component (see, for
example, Refs.~\onlinecite{Kroemer1985a,PhysRevB.84.045317}).
From the heterojunctions we have studied, three closed chains (contours) can be
formed in which the initial and the terminal components are the same:
SrTiO$_3$/PbTiO$_3$/BaTiO$_3$/SrTiO$_3$,
BaTiO$_3$/PbTiO$_3$/PbZrO$_3$/BaZrO$_3$/BaTiO$_3$, and
SrTiO$_3$/PbTiO$_3$/PbZrO$_3$/BaZrO$_3$/BaTiO$_3$/SrTiO$_3$.
The positions of the valence bands calculated from the obtained $\Delta E_v$
values for one of these chains is shown in Fig.~\ref{fig5}. When traversing
the contour, we never get zero (see the figure), the deviation is from
$-$0.027 to +0.539~eV. This means that the transitivity rule is not
applicable. The reason for this behavior is that in heterojunctions, the band
offsets actually depend on the in-plane lattice
parameter.~\cite{PhysRevB.34.5621, PhysRevB.44.5572,SolidStateCommun.124.63,
PhysRevB.67.155327}  If the lattice parameter in all the members of the
chain would be the same, then traversing the contour would result in zero
$E_v$ shift.%
	\footnote{This was indeed observed for three almost lattice-matched
	heterojunction pairs, Ge/GaAs, Ge/ZnSe, and
	GaAs/ZnSe.~\cite{Kroemer1985a}}
However, because all heterojunctions in the above chains have different
lattice parameters, the result is nonzero. Thus, in general, the transitivity
rule is untenable, and the corresponding error can exceed 0.5~eV.

\section{Conclusion}

In this work, the band offsets for eight heterojunctions between titanates
and zirconates of calcium, strontium, barium, and lead as well as for the
KNbO$_3$/NaNbO$_3$ heterojunction with the cubic perovskite structure were
calculated
from first principles. The effect of strain in contacting oxides on their
energy structure, many-body corrections to the position of the band edges
(calculated in the $GW$ approximation), and the splitting of the conduction
band resulting from spin-orbit interaction were consistently taken into
account. It was shown that the neglect of the many-body effects can lead to
errors in determination of the band offsets up to 0.36~eV. The failure of
the transitivity rule, which is often used to determine the band offsets in
heterojunctions, was demonstrated and its cause was explained.

\begin{acknowledgments}
This work was partially supported by the Russian Foundation for Basic Research
grant No.~13-02-00724. The calculations were performed on the laboratory
computer cluster and the ``Chebyshev'' supercomputer at Moscow State University.
\end{acknowledgments}


%
\providecommand{\BIBYu}{Yu}

\end{document}